\documentclass[a4,12pt,elsart]{article}
\usepackage{graphics}
\usepackage{epsfig}
\usepackage{epsf}

\begin{document}
\title{Hiking the hypercube: producers and consumers}

\author{Tanya Ara\'{u}jo\\
Research Unit on Complexity in Economics \\
and ISEG, Universidade T\'{e}cnica de Lisboa \\
{\small     {\em email}:tanya@iseg.utl.pt}\\
 G{\'e}rard~Weisbuch$^{*}$ \\
$^{*}$Laboratoire de Physique Statistique\footnote{
Laboratoire associ{\'e} au CNRS (UMR 8550), {\`a} l'ENS et
 aux Universit{\'e}s Paris 6 et Paris 7}
\\
     de l'Ecole Normale Sup{\'e}rieure, \\
    24 rue Lhomond, F-75231 Paris Cedex 5, France. \\
  {\small     {\em email}:weisbuch@lps.ens.fr}\\
 }
\maketitle

\abstract{We study the dynamics of co-evolution of producers
and customers described by bit-strings representing individual traits.
Individual ''size-like'' properties are controlled by
binary encounters which outcome depends upon a recognition process.
Depending upon the parameter set-up, mutual selection
 of producers and customers results in different 
types of attractors, either an exclusive niches 
regime or a competition regime.}

PACS: 89.65.-s  Social and economic systems

\section{Introduction: hikers of the hypercube}
\label{intro}


   The study of dynamical systems most often concerns
systems with few degrees of freedom, or spatial
systems. Physicists are less concerned with other
high dimensional systems. But many models
in biology and the social sciences are based 
on bit strings of often large dimensions.

  If we take the case of biology and especially
genomics, the genome or the succession of
monomers in biopolymers, is often represented by 
a bit string. This is e.g. the case for models 
of the origin of life \cite{ander}, immunology
 \cite{perel}, and of co-evolution \cite{coev}.

  These models are generally based on the following 
functional scheme:

\begin{displaymath}
  Genome \Longrightarrow recognition \Longrightarrow action \Longrightarrow
  results \Longrightarrow population \quad change.
\end{displaymath}
 
  The ''genome'' described by a bit string undergoes
selective encounters with other genomes and
some action with more or less successful results
influences a death and/or ''birth'' process.

These models also inspired social sciences:
Cultural and opinion dynamics have been described by
 \cite{culture,meet}. The minority game  \cite{minority}
applies to finance. Finally the genetic algorithm
approach \cite{hol} has a wide range of applications
ranging from co-evolution to optimisation.

  All these algorithms have in common the bit-string description
and most often a recognition process, but they differ 
by the consequences of significant encounters,
as stressed by the well known title in psychology:
''What do you do after your say Hello?''\cite{berne} .

  The present model is built along the above lines.
Let us consider a set of consumers possibly interested 
 in different products which characteristics are described
by bit strings. In the case of a car these characteristics
could be comfort, speed, size, gas consumption etc. 
We here simplify by considering binary, independent characteristics.

  The preferences of consumers $i$ depend upon 
the overlap of their own ideal string, their
 need string $S_i$, with the product set of characteristics $S_j$:
 \begin{equation} 
\begin{array}{lll} 
q_{ij} = \sum_{l=1}^k S^l_i .EQ. S^l_j 
\end{array}%
\end{equation} 
   Where the symbol $EQ$ stands for the logical
equivalence relation, which gives 1 if the two bits
are equal and 0 otherwise. The recognition condition
is that the overlap is larger than a fixed threshold.
Encounters and generated profits (or losses) influence
an integrated continuous variable which control
the survival of the agents. Our model, to be completely
defined in the next section, then defines
 a co-evolution dynamics of consumers and producers.

We are interested in the dynamics of population of agents on 
the hypercube and in the stable patterns
that may arise. We want to know how  these patterns
depend upon the parameters of the model: what are the
different regimes, where are their transitions.

  The model is based on a consumers/producers co-evolution (as in \cite{rui})
but it can also be applied to other domains.
In political science for instance, one could model
 the co-evolution of political parties platforms
and voters choices
as described by bit strings; each bit is now
the position of the party (or the voter) on some specific issue,
Europe, retirement policy, environment etc.

Other applications can be very concrete.
A major car constructor can in principle provide
a lot of options: their combination would correspond to more than 
ten thousands
 different cars. But not all combinations would sell well;
furthermore each change on the production chain has a cost.
The issue for the producers is which combination of options
should I propose to the public in view of its distribution
of preferences. The same issue is faced by the local dealer:
which stock of combinations should I keep 
available to the local public? 
 
 

\section{The Model} 
 
In the model there are $N_p$ producer agents and 
$N_c$ consumer agents. Each consumer prefers a product 
which characteristics are coded by a ''need''
string of $k$ bits. Each producer manufactures a single product 
which characteristics are also coded by a product
string of $k$ bits. The bit string of a consumer represents what the 
consumer agent {\it needs} and the 
bit string of a producer is a code for the {\it product} that he is 
able to supply.
 
In addition to the two bit strings that code for needs and products, 
the ''wealth'' of each agent is defined by a scalar variable $S$ or $C$,
 depending upon the agent 
type (consumer or producer, respectively). The variable $S_i$ 
represents the degree of satisfaction of the consumer $i$ and $C_j$ 
represents the capital accumulated by producer $j$.

In economy this role is played by money, but in other contexts 
it could be power or status. 
 
The dynamics of the model is described by:

 A {\bf recognition and transaction process}.

At each time step, all consumers look for the producer which product
is closest to their need (the product string with the largest overlap $q_{ij}$;
in the case of equality among several producers a random choice
 is made among them).
If the relative overlap $\frac{q_{ij}}{k}$ between the producer and the consumer
strings is
larger than a threshold $\theta$, a transaction occurs and the consumer
satisfaction is changed according to:
\begin{equation} 
\begin{array}{lll} 
S_{i}(t+1)=S_{i}(t)-a_c+\frac{q_{ij}}{k} \label{1.00} &  & 
\end{array}%
\end{equation} 
  Satisfaction is increased according the relative overlap
of the need and product strings. It is decreased by a constant 
cost per transaction $a_c$. 

 The equivalent updating of the producers capital is:
 \begin{equation} 
\begin{array}{lll} 
C_{j}(t+1)=C_{j}(t)-a_p+\sum_{i(j)}\frac{q_{ij}}{k} 
\label{2.00} & & 
\end{array}%
\end{equation} 
  Capital is increased by the set of transactions in which
producer $j$ was involved during the time interval, according the relative overlaps
of the need and product strings. The index $i(j)$
 runs over all the consumers $i$ that were supplied by 
producer $j$.
 Capital is decreased by a constant 
production cost $a_p$.

A  {\bf death and renewal process}.

  There is no upper limit to consumer satisfaction
or to producer capital. But due to the costs terms,
they may decrease to zero.

 A producer which capital
would become negative is destroyed and not renewed.

On the opposite, a consumer which satisfaction
would become negative disappears and is replaced 
by a new consumer, with a constant initial satisfaction
$S_0$ and a randomly generated need string.
 
\section{ Checking the parameter space }

 In principle there are eight independent parameters:

 \begin{itemize}
 \item The string length $k$ and the threshold $\theta$ for interaction.
 \item The consumption constants $a_p$ and $a_c$.
\item The initial numbers of producers ($N_p$) and consumers ($N_c$).
 \item The initial endowments of producers ($C_p$) and consumers ($S_c$).
 \end{itemize}

  In practice, because of the irreversibily of the
producers death process, $C_p$ and $N_p$ influence
mostly the early stages of the dynamics.

In terms of dynamical regimes and transitions, we might expect that
 the most important parameters are $a_c$ and $\theta$
( since we keep $k$ constant).

  In equation 1, the positive term $\frac{\theta}{k}$ is between 0 and 1.
Then, for large values of $a_c$, $a_c$ larger than 1, the change in consumer
satisfaction is negative: the dynamics is characterised by a renewal process
of consumers. Consumers are created with $S=S_c$, and their satisfaction 
can only linearly decrease to zero when they die. On average their 
satisfaction is  $S=S_c/2$. $a_c=1$ is then the critical value of
the satisfaction parameter 
separating the regime where some consumers may increase their satisfaction 
with time and survive, from a renewal regime when all consumers are condemned
to a rapid death.

  The threshold for interaction $\theta$ fixes a basin of satisfaction
in the neighborhood of the producer, which size is given by the number of sites
$n_s(\theta)$ which Hamming distance to its product string is less or equal
to $k-\theta$.
 Any consumer which string is in the basin of satisfaction of a producer may interact with her.
 For $k=10$, $\theta=1$ gives a 
neighbourhood of size 1, size 11 for  $\theta=0.9$, size 56 for $\theta=0.8$ etc.
Comparing $n_s(\theta) \times N_p$ to $2^k=1024$, the number of sites
on the hypercube, gives an approximation
of the condition in threshold separating the competition regime at 
low   $\theta$ values, from the ''niche'' regime. In the competition
regime, customers may choose between different 
producers in their neighborhood;
producers then competes for customers. In the niche regime,
they are so far apart that they don't compete.

 In the competition regime, according to $a_p$ values,
we might predict that competition will reduce a too large initial
 number of producers by bankruptcy, until they get closer to the critical
number where the niche regime is established.
But according to initial conditions, essentially $C_p$ and $N_p$,
the selection of surviving producers may end up in a
variety of attractors as shown by simulations.
Anyway, a maximum number of surviving producers 
in the competition regime can be estimated from equation 3. At equilibrium:
   \begin{equation} 
\begin{array}{lll} 
 a_p=\frac{N_c}{N_p} \frac{<q_{ij}>}{k} \\
 N_p =\frac{N_c}{a_p} \frac{<q_{ij}>}{k}
\label{4.00} & & 
\end{array}%
\end{equation} 
  These equations are based on an approximation of 
uniform distribution of product strings on the hypercube.
On average each producer is surrounded by  $\frac{N_c}{N_p}$
customers which contribute to its capital balance.
We will further observed that this upper bound is seldom reached.


\section{ Simulation results}

  Most simulations were made for a fixed set of parameters:
String length $k=10$,
 producers cost constant $a_p=4.5$,
number of consumers ($N_c$=1000) and 
 initial endowments of producers ($C_p=200$)
and consumers ($S_c=5$).
Simulation times were usually 2000, but we checked that
the stationary regime was reached by increasing simulation times
up to 10000.

  We studied the influence of consumers constant $a_c$,
threshold $\theta$ and $N_p$ the initial number of producers.

\subsection{Time evolution}

  Let us first monitor the time evolution of
 producers capital $C_i$, of the average consumer satisfaction $S$,
of the number of consumers death per time unit and of the 
surviving number of producers.

\begin{figure}[htbp]
\centerline{\epsfxsize=120mm\epsfbox{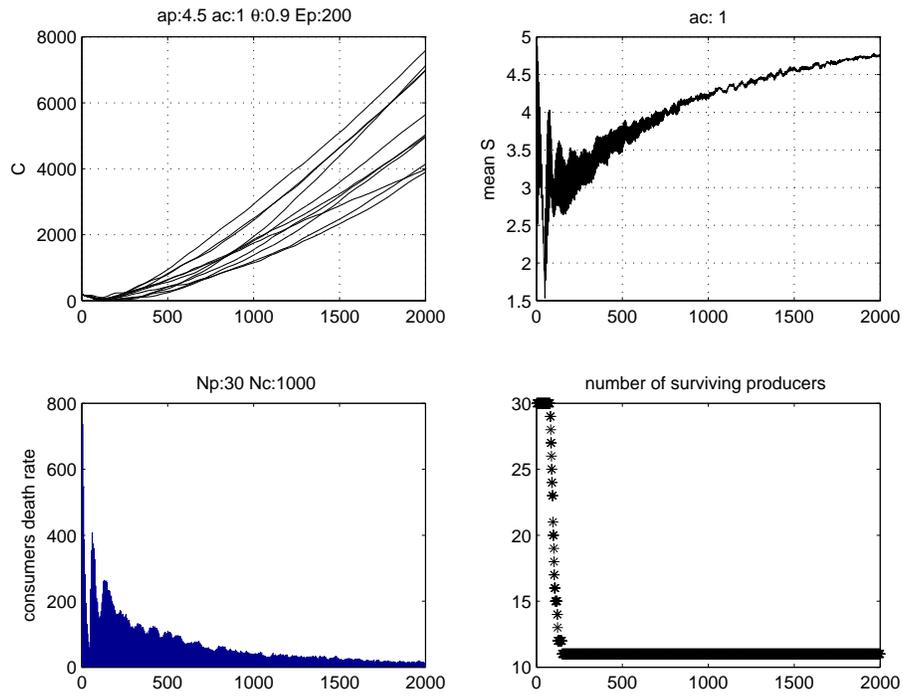}}
\caption{ Time evolution of
 producers capital $C_i$, of the average consumer satisfaction $S$,
of the number of consumer death per time unit and the number of surviving 
producers.
$N_p$ initial was 30, $N_c=1000$, $a_c=1$, $C_p=200$, $\theta=.9$. 
} 
\end{figure}
 On figure 1, where $N_p$ initial was 30, $N_c=1000$, $a_p=1$, $\theta=.9$,
we see on the lower right curve 
that producers are first selected: there number is decreased from 30
to 11, and then business becomes profitable ($C$ evolution, on the upper left
curves). The early producer selection gives figures much 
lower that the upper bound computed by equation 3 (which should be
around 200), but surviving producers make profit.

Consumers average satisfaction is always lower than 5: since
$a_c=1$ no consumer
ever increase satisfaction alive, but the random generation of
reborn consumers allow some of them, ''the condensed'' consumers,
to reach the cell of one producer
where they may survive for ever. Average $S$ reflects the division
between a fraction $f$ of condensed consumers with $S=5$
and of starving consumers with average $S=2.5$. $f$ then obeys:
\begin{equation}
  f \times 5 + (1-f) \times 2.5 = <S>
\end{equation}
  
 \begin{equation}
  f  = \frac{<S>-2.5}{2.5}
\end{equation}
 The increase of $<S>$ towards 5 with time then reflects the increase of the
fraction of ''condensed'' consumers towards 1. This effect is confirmed by 
decrease of the death rate on the lower left plot.

\begin{figure}[htbp]
\centerline{\epsfxsize=120mm\epsfbox{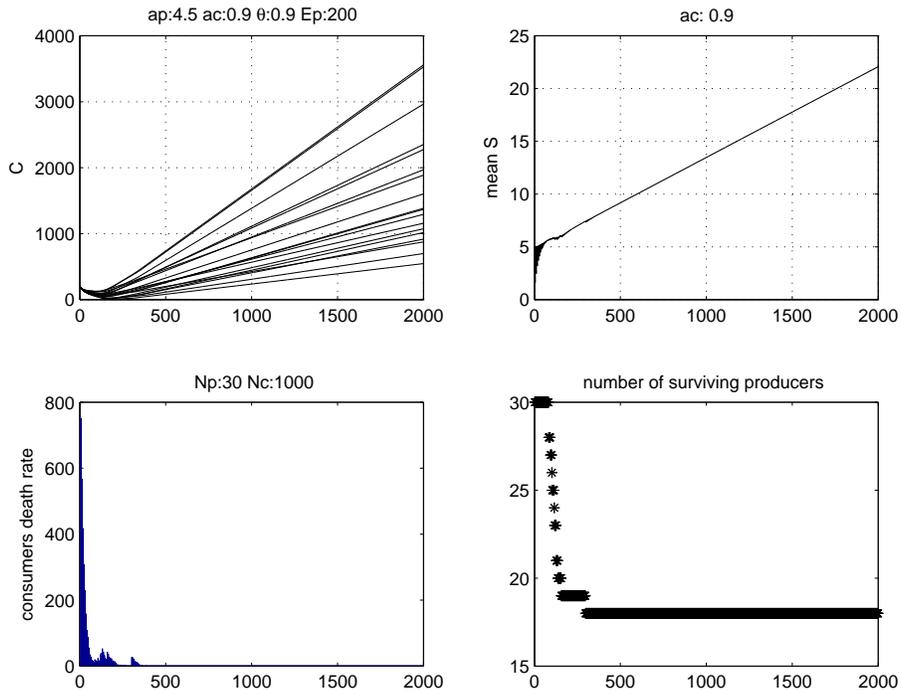}}
\caption{ Time evolution of
 producers capital $C_i$, of the average consumer satisfaction $S$,
of the number of consumer death per time unit and of 
the number of surviving  producers strings.
$N_p$ initial was 30,$N_c=1000$, $\theta=.9$, $C_p=200$, as on the previous figure
but the product cost is now  $a_p=0.9$ instead of 1.
} 
\end{figure}

  With $a_p=0.9$, smaller than one, those condensed consumers 
in the neighborhood of the producers can even increase satisfaction
during their infinite lifetime as observed on the right upper
corner. Because of the lower stress on consumers, they early find spots
 where they are able to survive, and they stabilize in the neighborhood
of producers much earlier in time. The diversity of producers is
higher: 17 survive.

   \subsection{  Attractors}
   
  Let us now do a systematic study of the attractors of the dynamics.
We checked the asymptotic state of the system through a similar set of variables:
producers capital $C_i$, average consumer satisfaction $<S>$,
 number of surviving producers.

  We also tried to characterise the degree of order
or of diversity of the attractor configuration
and used the overlap of buyers need strings
as the order function. This is the  same
notion as used in spin glass theory. We
 then checked the fraction $\frac{h(10)}{h(5)}$
of the histogram of overlaps among buyers need strings. 
This gives us an indication about how many consumers are condensed
 and what is their repartition (even or uneven)
in the neighborhood of producers.

In the next four figures,
the horizontal variable is $N_p$ the initial number of producers.
The four set of figures correspond to two values of $\theta$, 09 and 0.8,
 and $a_p$, 0.9 and 1. These results were averaged over 100
different random samplings of initial configurations.

   There is not much to add about the influence $a_p$ from these figures.

  But the influence of $\theta$ on the transition between a competition
regime ($\theta=.8$) and a niche regime ($\theta=.9$)
 is now made clear. 

In the niche regime, $\theta=.9$,
few producers survive, even for large initial number of producers
(figure 3).  Their number saturates around 20 
for higher values of the initial producer number.
Producers make a lot of profit (large $C$ values) (figure 4).

Consumers satisfaction depends upon $a_c$
 (figure 5).

 Significant values of $h(1)$
are observed when $a_c$ and $\theta$
are both large.


\begin{figure}[htbp]
\centerline{\epsfxsize=120mm\epsfbox{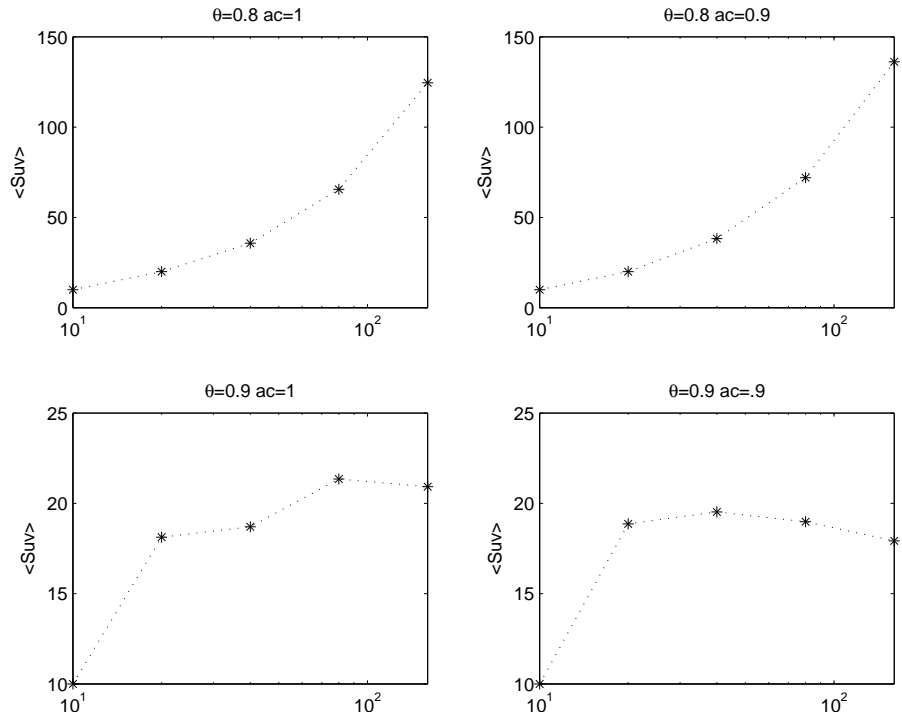}}
\caption{ 
 Number of surviving producers. These are taken at time=2000.
$N_p$ is varied along the x axis (logarithmic scale).
 Each of the four plots correspond to a given pair of  $a_p$
and $\theta$ values, upper plots to a threshold $\theta=0.8$
lower  plots to a threshold $\theta=0.9$, left plots to 
consumption $a_c=1$ and right plots to 
consumption $a_c=0.9$.
} 
\end{figure}


\begin{figure}[htbp]
\centerline{\epsfxsize=120mm\epsfbox{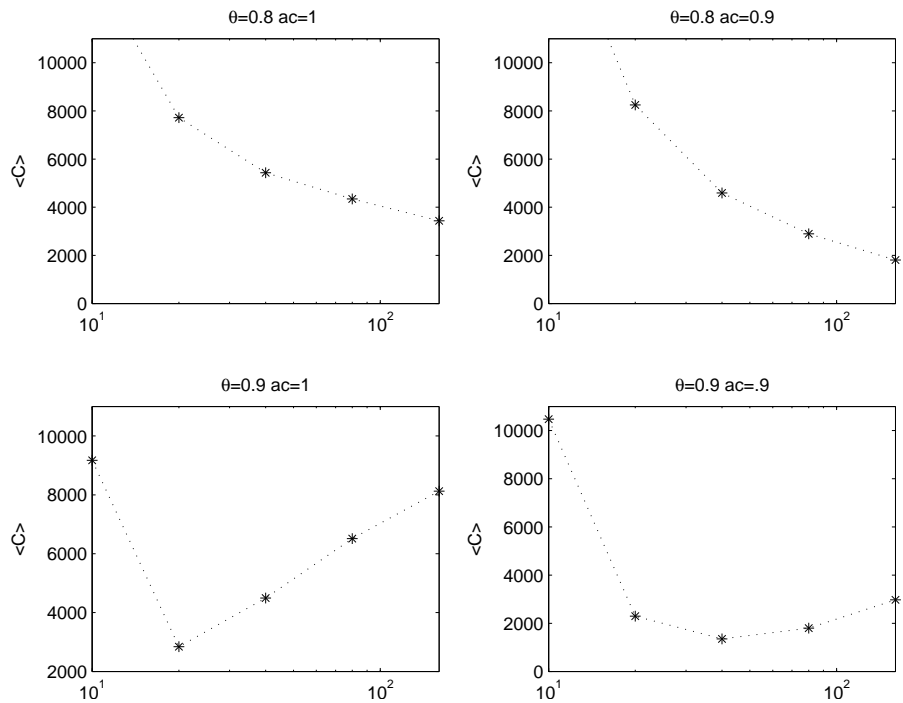}}
\caption{  
 Averaged producer capital $C_i$ at time=2000.
 $N_p$ is varied along the x axis (logarithmic scale).
 Each of the four plots correspond to a given pair of  $a_p$
and $\theta$ values.
} 
\end{figure}

\begin{figure}[htbp]
\centerline{\epsfxsize=120mm\epsfbox{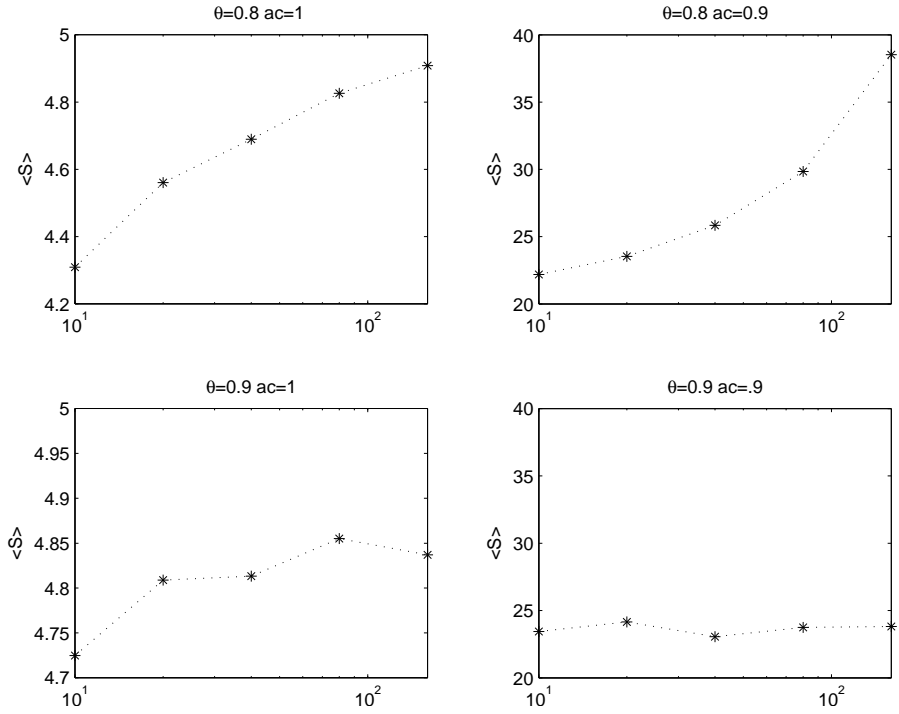}}
\caption{  Average consumer satisfaction $S$ at time=2000.
$N_p$ is varied along the x axis (logarithmic scale).
 Each of the four plots correspond to a given pair of  $a_c$
and $\theta$ values, upper plots to a threshold $\theta=0.8$
lower  plots to a threshold $\theta=0.9$, left plots to 
consumption $a_c=1$ and right plots to 
consumption $a_c=0.9$.
} 
\end{figure}

\begin{figure}[htbp]
\centerline{\epsfxsize=120mm\epsfbox{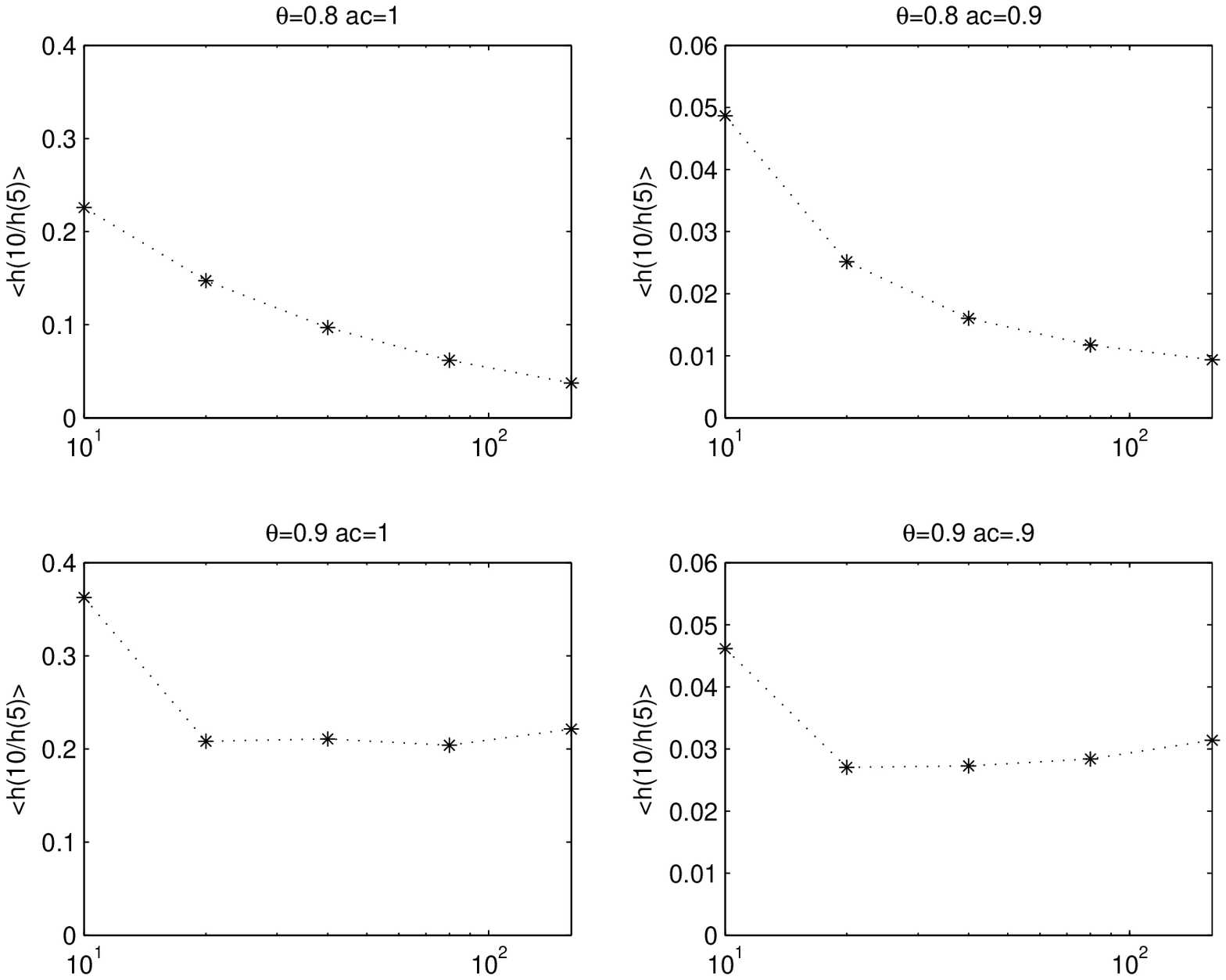}}
\caption{ Relative size of the condensed fraction ( $\frac{h(10)}{h(5)}$)
of consumers at time=2000.
$N_p$ is varied along the x axis (logarithmic scale).
} 
\end{figure}

  On figure 6, the large value of the 10 overlap bin
observed when $a_c=1$  and $\theta=.9$ is an indication that 
many customers have the same need string. This concentration  
of need strings, co-occuring with small number of surviving 
producers, is obviously due to their ''condensation''
on producers ''product'' strings. Furthermore, the large value
of 0.2 observed for $a_c=1$ and $\theta=0.9$ tells
us that their distribution among different producers 
is uneven; an even distribution would correspond
to the inverse number of surviving producers,
0.05.

  The observation of the time evolution of producer capital,
of their number, and similar observations
on consumers clearly demonstrate two phases of the dynamics
at least when $a_c<1$ and $\theta$ is sufficiently large. 

\begin{itemize}
\item An initial transient period, during which those 
customers which are not close enough to producers
die. An equivalent selection process occurs for
producers which are selected against when 
customers are not dense enough in their neighborhood to support them.

\item A stationary period, following the selection period.
During this period, customers get enough supply
and producer capital increase. Depending upon consumers
cost $a_c$, their satisfaction may increase when $a_p<1$
or saturate when $a_c=1$. Depending upon the threshold 
for recognition $\theta$, two distinct stationary regimes
are observed: 
\begin{itemize}
\item A niche regime when customers condensate
close to producers, at lower producer density and higher threshold
values.
\item A competition regime when producers compete for 
customers, at higher producer density and lower threshold
values.
\end{itemize}
\end{itemize}

\subsection{ Patterns on the hypercube}

   Further observations show that the selection process
does not give rise to uniform densities of customers and producers
on the hypercube: this process is spatially unstable. When a region is depleted 
say in customers, the density of producers is also depleted,
and a further depletion of customers also occurs.
As a result the selection process goes much further in reducing the number
of producers that equation 3 would suggest;
surviving producers then make profit and the histogram of their
Hamming distance  is biased towards the bins lower than 5
(which would correspond to a random distribution),
 as observed on figure 7. This figure which
represent the result of 100 simulations was taken
under severe constraints, $a_p=10$ and $C_p=1000$, but the number of survivors
is much less than the prediction of equation 3 (100).
Producer strings concentrate in a small region of the hypercube,
as observe on the histogram of distances, and by direct examination
of the product strings (not shown here). 
 
\begin{figure}[htbp]
\centerline{\epsfxsize=120mm\epsfbox{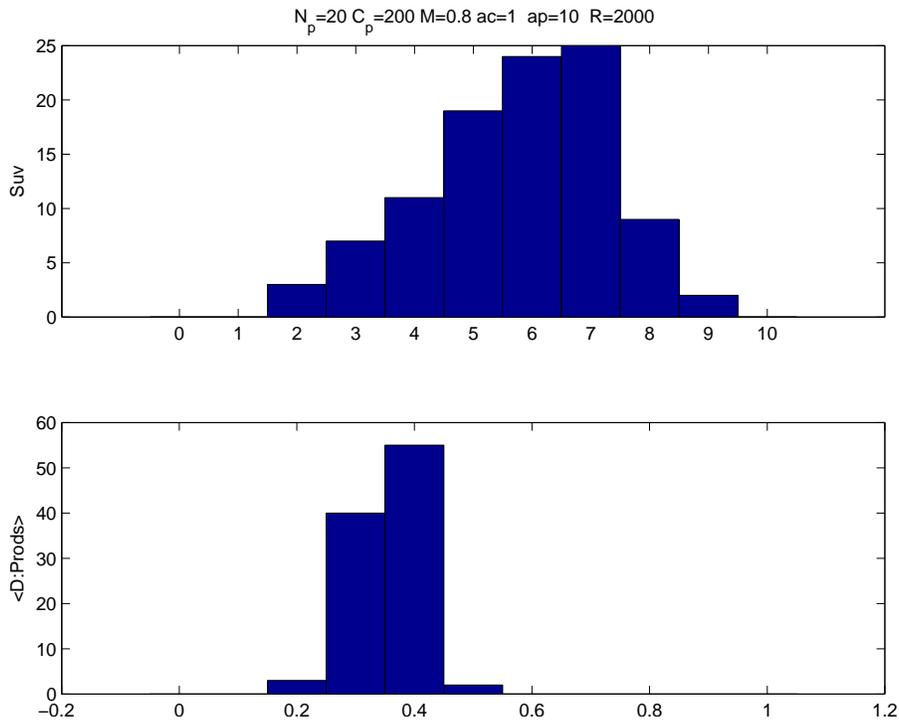}}
\caption{ Instability of the producer selection process.
The histogram of the number of surviving producers,
the upper diagram, displays a strong reduction from initial
producer number,
$N_p=100$, with large deviations. The lower histogram displays
distances among producer strings, and shows that they are
concentrated in only a small part the hypercube.
  } 
\end{figure}

The initial dynamics 
of accumulation of customers in the
satisfaction basin of the survivors displayed on figure 8
comfort
our hypotheses. One observes a strong population increase
in the satisfaction basin of the
surviving producers and a slow decay
in the neighborhood of the opposite strings.
The random initial population of any
2-neighbourhood is 56 on average. 

The two growth regimes, fast initial increase
and slower later increase are understandable:
the initial population increase (during the first 100 time steps)
is due to the fact that any site on the 2-neighbourhood
of a surviver has a much longer  lifetime
than those in the desert (the desert time life=5). Close to the 
producers the life time is infinite at 0 distance,
50 at distance 1,  25 at distance 2. Re-born consumers
(1/5) have a 1000/56 chance to get to an attraction
basin of a surviver, which gives the observed figure of 100 time steps
for the steep increase duration.
   Later the slow growth correspond
to the replacement of consumers in the attraction
basin by those who hit the surviver site and thus eternity.

\begin{figure}[htbp]
\centerline{\epsfxsize=120mm\epsfbox{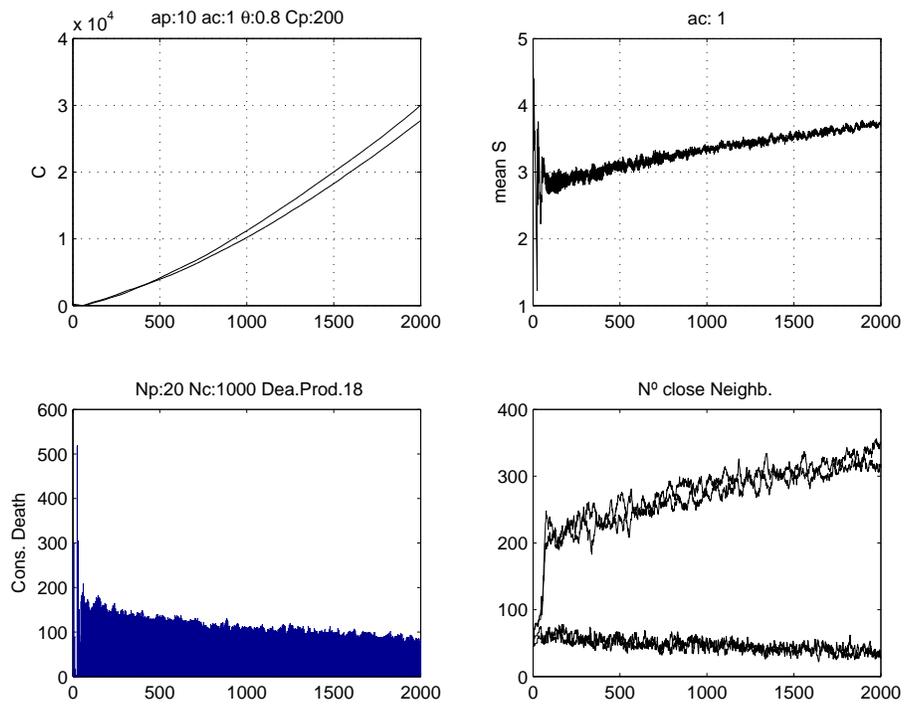}}
\caption{  Time evolution of
 producers capital $C_i$, of the average consumer satisfaction $S$,
of the number of consumer death per time unit and of the number of customers
in the attraction basin of the two surviving producers (rising curves)
and of two random sites.
$N_p$ initial was 20,$N_c=1000$, $a_c=1$, $\theta=1$.
But the value of $a_p=10$ gives rise to a strong selection of producers. 
  } 
\end{figure}

\section{ Conclusions}

   The present version of the model is evidently very
crude with respect to the more general aspects than 
one would like to investigate in this framework.
  One would like to better understand the dynamics
of co-evolution when producers are endowed
with stategic behaviour. This line of research 
has already been investigated by \cite{rui}.
We can predict from our present study that
producers strategies may be influential 
and provide higher gains in the competition
regime. When the  niche regime is established 
their efficiency will be extremely limited because
of the condensation of consumers in the neighbourhood of
producers. Other renewal processes can be imagined
for consumers such as myopic learning or reproduction 
with noise (with probably equivalent results in the
two latter cases).

  In other words, the present study has set the stage
for more intricate studies in co-evolution.  

Acknowledgments: The first author (TA) thanks R. Vilela Mendes for
the illuminating discussions on the original model. The present work
was started on the occasion of G\'{e}rard Weisbuch   visit
to Lisbon,
supported by FCT-Portugal, grant PDCT/EGE/60193/2004. GW was also
supported by E2C2 NEST 012410 EC grant.

\end{document}